\documentclass[aps,prapplied,reprint,superscriptaddress]{revtex4-2}
\usepackage{graphicx}
\usepackage[caption = false]{subfig}
\usepackage{url}\usepackage{hyperref}
\usepackage{braket}
\usepackage{amsmath}

\newcommand{\ad}{{a^\dagger}}
\newcommand{\ads}[1]{{a_{#1}^\dagger}}
\newcommand{\us}{\mathrm{\mu s}}
\newcommand{\idx}{n}
\newcommand{\idy}{m}
\newcommand{\state}[1]{\ket{\psi_{#1}}}
\newcommand{\excitation}[2]{{\Delta\omega_{#1,#2}}}
\newcommand{\Deltain}{{\Delta_\mathrm{in}}}
\newcommand{\Omegain}{{\Omega_\mathrm{in}}}
\newcommand{\thetain}{{\theta_\mathrm{in}}}
\newcommand{\omegar}{{\omega_\mathrm{r}}}
\newcommand{\Deltar}{{\Delta_\mathrm{r}}}
\newcommand{\omegap}{{\omega_\mathrm{p}}}
\newcommand{\Deltap}{{\Delta_\mathrm{p}}}
\newcommand{\thetap}{{\theta_\mathrm{p}}}
\newcommand{\Pp}{{P_\mathrm{p}}}

\begin{document}
	\title{Change in bit-flip times of Kerr parametric oscillators caused by their interactions}

\author{Yuya Kano}
\author{Yohei Kawakami}
\affiliation{Secure System Platform Research Laboratories, NEC Corporation, Kawasaki, Kanagawa 211-0011, Japan}
\affiliation{NEC-AIST Quantum Technology Cooperative Research Laboratory, National Institute of Advanced Industrial Science and Technology (AIST), Tsukuba, Ibaraki 305-8568, Japan}

\author{Shumpei Masuda}
\affiliation{NEC-AIST Quantum Technology Cooperative Research Laboratory, National Institute of Advanced Industrial Science and Technology (AIST), Tsukuba, Ibaraki 305-8568, Japan}
\affiliation{National Institute of Advanced Industrial Science and Technology (AIST), Tsukuba, Ibaraki 305-8565, Japan}
\affiliation{Global Research and Development Center for Business by Quantum-AI technology (G-QuAT), National Institute of Advanced Industrial Science and Technology (AIST), Tsukuba, Ibaraki 305-8560, Japan}

\author{Tomohiro Yamaji}
\author{Aiko Yamaguchi}
\affiliation{Secure System Platform Research Laboratories, NEC Corporation, Kawasaki, Kanagawa 211-0011, Japan}
\affiliation{NEC-AIST Quantum Technology Cooperative Research Laboratory, National Institute of Advanced Industrial Science and Technology (AIST), Tsukuba, Ibaraki 305-8568, Japan}

\author{Tetsuro Satoh}
\affiliation{Global Research and Development Center for Business by Quantum-AI technology (G-QuAT), National Institute of Advanced Industrial Science and Technology (AIST), Tsukuba, Ibaraki 305-8560, Japan}

\author{Ayuka Morioka}
\author{Kiyotaka Endo}
\author{Yuichi Igarashi}
\author{Masayuki Shirane}
\affiliation{Secure System Platform Research Laboratories, NEC Corporation, Kawasaki, Kanagawa 211-0011, Japan}
\affiliation{NEC-AIST Quantum Technology Cooperative Research Laboratory, National Institute of Advanced Industrial Science and Technology (AIST), Tsukuba, Ibaraki 305-8568, Japan}

\author{Tsuyoshi Yamamoto}
\affiliation{Secure System Platform Research Laboratories, NEC Corporation, Kawasaki, Kanagawa 211-0011, Japan}
\affiliation{Global Research and Development Center for Business by Quantum-AI technology (G-QuAT), National Institute of Advanced Industrial Science and Technology (AIST), Tsukuba, Ibaraki 305-8560, Japan}

\date{\today}

\begin{abstract}
We experimentally investigate how interactions between Kerr parametric oscillators (KPOs) degrade their bit-flip times, where a bit flip is defined as a transition between the two degenerate ground states of a KPO.
Interactions between KPOs cause quantum states of KPOs to leak outside the computational subspace, leading to bit flips.
Bit flips degrade fidelity and pose a significant problem for KPO-based quantum information processing.
We performed an experiment in which a weak microwave signal is injected into one KPO to emulate photon injection from another KPO, and find that the bit-flip time decreases by an order of magnitude due to induced excitations, depending on the frequency and power of the injected signal.
Methods to mitigate the decrease in bit-flip times caused by interactions between KPOs are discussed, including adjusting the pump frequencies, coherent-state amplitudes, and couplings between KPOs.
These findings provide valuable insights for scaling up KPO-based quantum computers.
\end{abstract}

\maketitle
	\section{Introduction}\label{sec:introduction}

A Kerr parametric oscillator (KPO) is a Kerr nonlinear resonator with a parametric drive~\cite{Goto_2016}.
KPOs have promising applications in quantum annealing~\cite{Goto_2016, Puri_2017, Nigg_2017, Zhao_2018, Onodera_2020, Kewming_2020}, universal quantum computation~\cite{Goto_2016b, Puri_2017b}, and other quantum information processing tasks~\cite{Goto_2018, Dykman_2018, Rota_2019}.
A key advantage of KPOs is their tolerance to bit flips caused by photon losses~\cite{Puri_2017}.
A bit flip is defined as a transition between the two degenerate ground states of a KPO, with the two coherent states $\ket{\pm\alpha}$ considered as the two computational states.

Realizing practical quantum computers with KPOs requires fabricating quantum chips with a large number of KPOs.
Quantum chips with two or four KPOs have been experimentally demonstrated~\cite{Yamaji_2023, Hoshi_2025, kawakami2025fourbodyinteractionskerrparametric}.

In this paper, we investigate how interactions between KPOs cause quantum states of KPOs to leak outside the computational subspace.
In a multi-KPO system, interactions between KPOs arise from e.g. capacitive couplings (either intentional by design or unintentional due to their physical proximity within the chip~\cite{Barends_2013}), or an effective coupling via a common transmission line~\cite{van_Loo_2013,Lalumi_re_2013,Gheeraert_2020} that can be used to implement frequency-multiplexed readout~\cite{Jerger_2011,Chen_2012,Jeffrey_2014} for KPOs.
These inter-KPO interactions facilitate photon exchange, which leads to leakage~\cite{Puri_2019} and thus to bit flips, as detailed in Sec.~\ref{sec:bitflip}.
Leakage degrades fidelity and poses a significant problem for KPOs, as it does for transmon qubits~\cite{Sank_2016}, especially in scaling up to larger systems.

To investigate the leakage of KPOs caused by their interactions, we performed an experiment in which we injected a weak microwave signal into one KPO to emulate photon injection from another KPO, and measured changes in the bit-flip time of the KPO.
The difference between probabilities of the two ground states decreases exponentially with time, and the bit-flip time is defined as the time constant of the exponential decay, similarly to Ref.~\cite{Su_2025}.
We find that the bit-flip time may decrease by an order of magnitude due to an induced excitation that leads to bit flips, and also that the decrease can be mitigated through adjustment of e.g. pump frequencies, coherent-state amplitudes, and couplings between KPOs.

This paper is organized as follows.
Section~\ref{sec:bitflip} describes the mechanism by which inter-KPO interactions induce leakage and bit flips.
Section~\ref{sec:device} details the design and fabrication of the KPO chip.
Section~\ref{sec:exp1b} presents the experimental result on bit-flip-time changes of a KPO under an injection of a weak microwave signal.
Section~\ref{sec:discussion} discusses methods to mitigate leakage and bit flips of KPOs.
	\section{Interactions between KPOs leading to bit flips}\label{sec:bitflip}
We first describe bit flips of a KPO and then explain how inter-KPO interactions induce them.
The Hamiltonian of a single KPO in a frame rotating at frequency $\omega_p/2$ is~\cite{Goto_2016}:
\begin{equation}
	\frac{H_\mathrm{1KPO}}{\hbar} = \Deltar \ad a + \frac{K}{2} \ad^2 a^2 + \frac{p}{2}(\ad^2+a^2). \label{eq:oneKPO}
\end{equation}
Here, $\omegap$ is the frequency of the injected parametric drive (pump), $\Deltar=\omegar-\frac{\omegap}{2}$ is the pump detuning, $\omegar$ is the resonance frequency of the KPO, $a$ is the photon annihilation operator, $K<0$ is the Kerr nonlinearity of the KPO, and $p$ is the pump amplitude.
When the pump amplitude is slowly increased for an adiabatic evolution of a KPO from a vacuum state, the KPO transitions to a superposition of two energy-degenerate coherent states $\ket{\pm\alpha}$ with amplitude $\alpha\simeq\sqrt{(p+\Deltar)/|K|}$~\cite{Goto_2016}, where the two coherent states correspond to the two stable points of the double-well metapotential of the KPO~\cite{Puri_2017}.
In this paper, we consider the two coherent states $\ket{\pm\alpha}$ as the computational states, e.g. $\ket{-\alpha}$ as state 0 and $\ket{\alpha}$ as state 1, which are used for quantum annealing~\cite{Goto_2016, Puri_2017}.
We define a bit flip as a transition of a KPO between the two computational states.

\begin{figure}
	\centering
	\includegraphics[width=.8\linewidth]{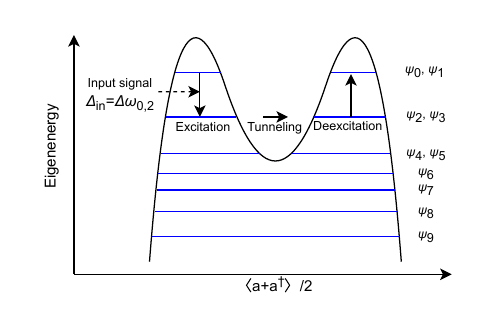}
	\caption{
		Schematic diagram of the metapotential of a KPO, together with the eigenenergies $\omega_\idx$ of the Hamiltonian eigenstates $\state{\idx}$, and the process of bit flip.
		Some higher excited states are not confined within the two wells.
		Eigenenergies are shown for a KPO with a coherent-state amplitude of $\alpha=2.8$, corresponding to the experiment in Sec.~\ref{sec:exp1b}.
	}
	\label{fig:bitflip}
\end{figure}

Several excited Hamiltonian eigenstates are also confined within the two wells of the KPO metapotential~\cite{Puri_2017,Frattini_2024}, as illustrated in Fig.~\ref{fig:bitflip}.
We denote Hamiltonian eigenstates as $\state{\idx}$, and they are sorted in a descending order of their eigenenergies $\omega_\idx$ considering the Kerr nonlinearity being negative.
The ground states $\state{0}$ and $\state{1}$ correspond to the two computational states when $\alpha$ is large and $\braket{\alpha|-\alpha}=0$~\cite{Puri_2019}, and in this paper we use $\state{0} = \ket{-\alpha} $ and $\state{1} = \ket{\alpha}$.

We now explain how inter-KPO interactions induce bit flips. 
Consider a two-KPO system with a two-body coupling. 
In a rotating frame defined by $\frac{H_0}{\hbar} = \sum_{\idx=1,2} \frac{\omegap_\idx}{2}\ads{\idx} a_\idx$, the Hamiltonian is~\cite{Hoshi_2025}:
\begin{multline}
	\frac{H_\mathrm{2KPO}(t)}{\hbar} = \\
	\sum_{\idx=1,2} \left[ \Deltar_\idx \ads{\idx} a_\idx + \frac{K_\idx}{2} {\ads{\idx}}^2 a_\idx^2 + \frac{p_\idx}{2}({\ads{\idx}}^2+a_\idx^2) \right] \\
	+ g \left[ e^{-i(\Deltap t + \thetap) } \ads{1} a_2 + e^{i(\Deltap t + \thetap)} a_1 \ads{2} \right]. \label{eq:twoKPOs}
\end{multline}
Here, $\idx $ $(=1,2)$ indexes the two KPOs, $\Deltar_\idx=\omegar_\idx-\frac{\omegap_\idx}{2}$ is the pump detuning, $g$ is the two-body coupling between the two KPOs, $\Deltap=(\omegap_2-\omegap_1)/2$ is half of the difference between the two pump frequencies, and $\thetap=({\thetap}_2-{\thetap}_1)/2$ is half of the difference between the two pump phases ${\thetap}_1$, ${\thetap}_2$.
Symbols $\omegap_\idx$, $a_\idx$, $\omegar_\idx$, $K_\idx$, and $p_\idx$ have the same meanings as those in Eq.~\eqref{eq:oneKPO}, but for each KPO.
The term proportional to $g$ in Eq.~\eqref{eq:twoKPOs} describes photon exchange between KPOs.
A photon injected into a KPO from another KPO can cause excitation, increasing the probability to tunnel between the two metapotential wells~\cite{Frattini_2024}.
This tunneling may be followed by a deexcitation due to photon loss that leads to a qubit state different from that before the excitation, resulting in a bit flip.
Thus, inter-KPO interactions cause the bit-flip time to decrease.

The discussion above can be understood intuitively by approximating the effect of a two-body coupling by an injection of a microwave signal to one KPO.
Under an approximation that KPO2 is in a coherent state $\ket{\alpha_2}$, we replace $a_2$ with $\alpha_2$ in Eq.~\eqref{eq:twoKPOs}.
The resulting effective Hamiltonian for KPO1 is (dropping constant terms):
\begin{multline}
	\frac{H_\mathrm{KPO1}(t)}{\hbar} = \Deltar_1 \ads{1} a_1 + \frac{K_1}{2} {\ads{1}}^2 a_1^2 + \frac{p_1}{2}({\ads{1}}^2+a_1^2) \\
	+ g \alpha_2 \left[ e^{-i(\Deltap t + \thetap)} \ads{1}  + e^{i(\Deltap t + \thetap)} a_1 \right]. \label{eq:approx}
\end{multline}
This matches the Hamiltonian of one KPO with an injected microwave signal as expressed below~\cite{Masuda_2021}, when $\Omega_\mathrm{in}= g\alpha_2$, $\Deltain=\Deltap$, and $\thetain=\thetap$.
\begin{multline}
	\frac{H_\mathrm{in}(t)}{\hbar} = \Deltar_1 \ads{1} a_1 + \frac{K_1}{2} {\ads{1}}^2 a_1^2 + \frac{p_1}{2}({\ads{1}}^2+a_1^2) \\
	+ \Omega_\mathrm{in} \left[ e^{-i(\Deltain t + \thetain)} \ads{1}  + e^{i(\Deltain t + \thetain)} a_1 \right]. \label{eq:input}
\end{multline}
Here, $\Omega_\mathrm{in}$ is the input-signal amplitude, $\Deltain=\omega_\mathrm{in}-{\omegap}_1/2$ is the input-signal detuning with $\omega_\mathrm{in}$ being the input-signal frequency, and $\thetain$ is the input-signal phase.
As shown in Fig.~\ref{fig:bitflip}, when $\Deltain = \excitation{\idx}{\idy} (\idx<\idy)$, where $\excitation{\idx}{\idy} \equiv \omega_\idy-\omega_\idx$ is the excitation energy between two Hamiltonian eigenstates, resonant excitation occurs and leads to bit flips.
	\section{Details of KPO chip}\label{sec:device}
	
\begin{figure}
	\centering
	\subfloat[]{\includegraphics[width=.8\linewidth]{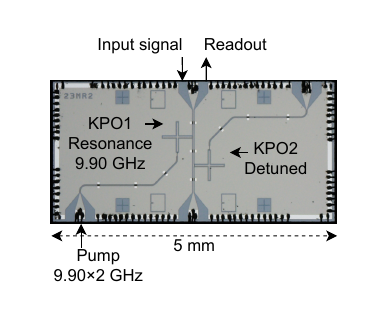}\label{fig:devicePhoto}} \\
	\subfloat[]{\includegraphics[width=\linewidth]{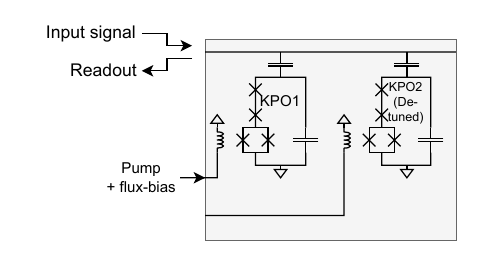}\label{fig:deviceSchematic}} \\
	\caption{
		(a) Optical micrograph of the KPO chip.
		Setup of the experiment in Sec.~\ref{sec:exp1b} is shown.
		(b) Schematic diagram of the KPO chip. 
        "X" symbols represent Josephson junctions.
	}
	\label{fig:device23mr2}
	\end{figure}

We fabricated a device with two Josephson parametric oscillators (JPOs)~\cite{Lin_2014} as shown in Fig.~\ref{fig:device23mr2}.
A JPO is a parametric oscillator whose nonlinearity is implemented using Josephson junctions, and it is called a KPO when operated in the single-photon Kerr regime~\cite{Kirchmair_2013}.
Each JPO comprises a superconducting quantum interference device (SQUID) and a capacitor, with two additional Josephson junctions in series with the SQUID to reduce the magnitude of the Kerr nonlinearity~\cite{Eichler_2014}.
A pump line is inductively coupled to the SQUID for applying a pump signal to induce the KPO oscillation and a flux-bias current to control the resonance frequency of the KPO.
The device is designed with a focus on frequency-multiplexed readout of two KPOs, with both capacitively coupled to a common transmission line for readout (readout line).
Single-shot measurement of the states of a KPO is performed by measuring its photons leaking into the readout line~\cite{Wang_2019, Yamaji_2022, Yamaji_2023, Masuda:2024djh}, as shown in Fig.~\ref{fig:deviceSchematic}.
The device is placed in a magnetic shield within a dilution refrigerator cooled to below 15~mK.

The device is fabricated using a Nb film with a thickness of 100~nm sputtered on a high-resistivity Si substrate with a thickness of 380~$\mathrm{\mu m}$.
The circuit is patterned with photolithography using dry etching with $\mathrm{CF_4}$ gas.
Josephson junctions are fabricated with an angled shadow evaporation of Al, with a prior removal of surface oxides on the Nb film by Ar-ion milling.
After the fabrication of Josephson junctions, airbridges are fabricated using an Al film with a method similar to that in Ref.~\cite{Yamaji_2025}.
	\section{Change in bit-flip time with respect to an injected microwave signal}\label{sec:exp1b}

As discussed in Sec.~\ref{sec:bitflip}, inter-KPO interactions degrade the bit-flip times of KPOs.
We test this effect experimentally based on Eq.~\eqref{eq:input} by injecting a weak microwave signal into a KPO and measuring the resulting change in its bit-flip time.
As shown in Fig.~\ref{fig:devicePhoto}, a pump signal induces the oscillation of KPO1, and an input signal is simultaneously injected.
Flux-bias current is applied to the SQUID loop of KPO1 to set its resonance frequency, while KPO2 is intentionally detuned (no flux bias applied) to a frequency above 11 GHz.
The details of the experimental setup are described in Appendix~\ref{sec:setup}.

\begin{figure}
	\centering
	\subfloat[]{ \includegraphics[width=.5\linewidth]{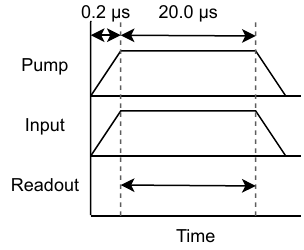}\label{fig:exp1bTiming} }
	\subfloat[]{\includegraphics[width=.5\linewidth]{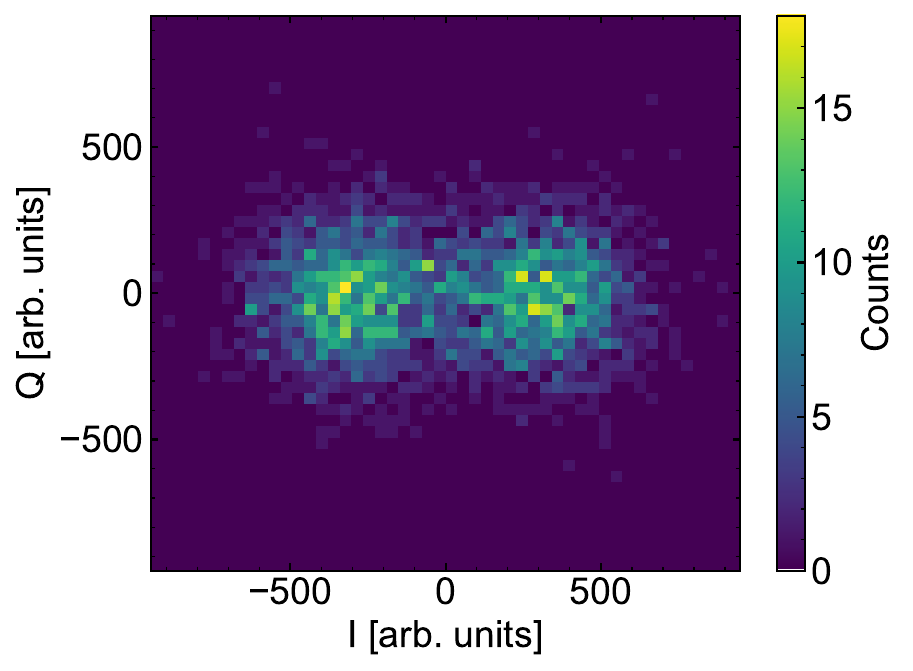}\label{fig:exp1bIQ} } \\
	\subfloat[]{\includegraphics[width=.5\linewidth]{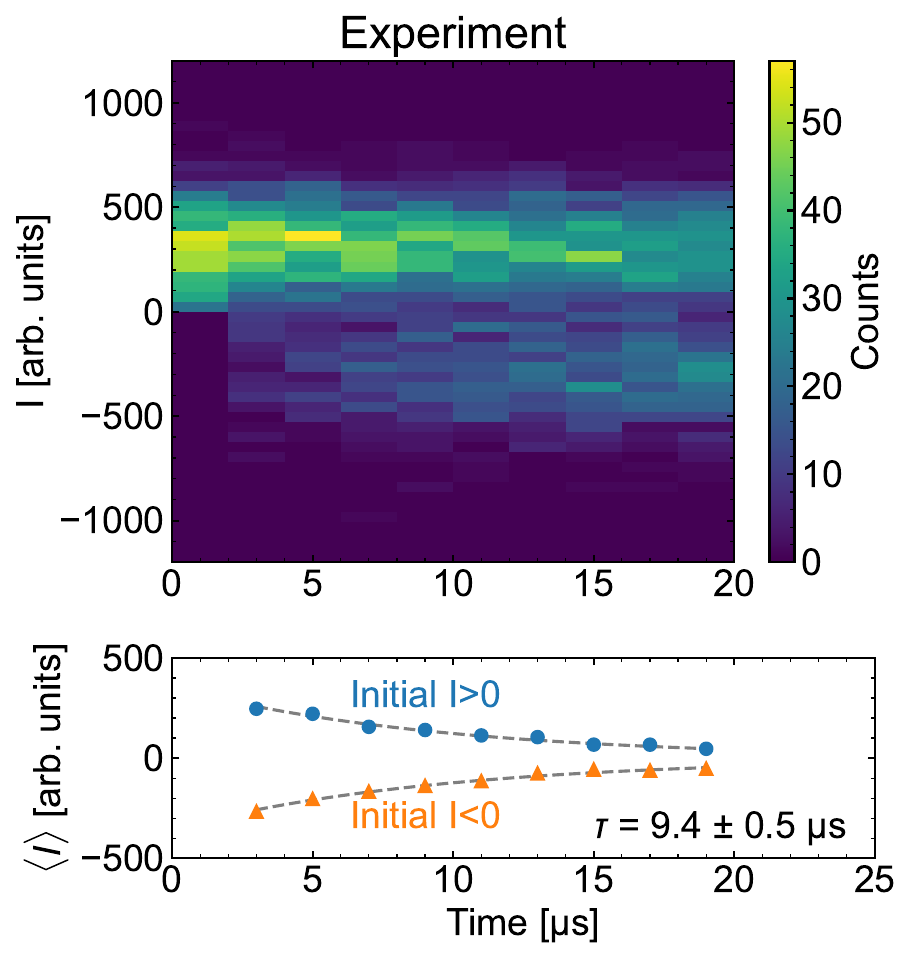}\label{fig:exp1bComparisonExpNoInput}}
	\subfloat[]{\includegraphics[width=.5\linewidth]{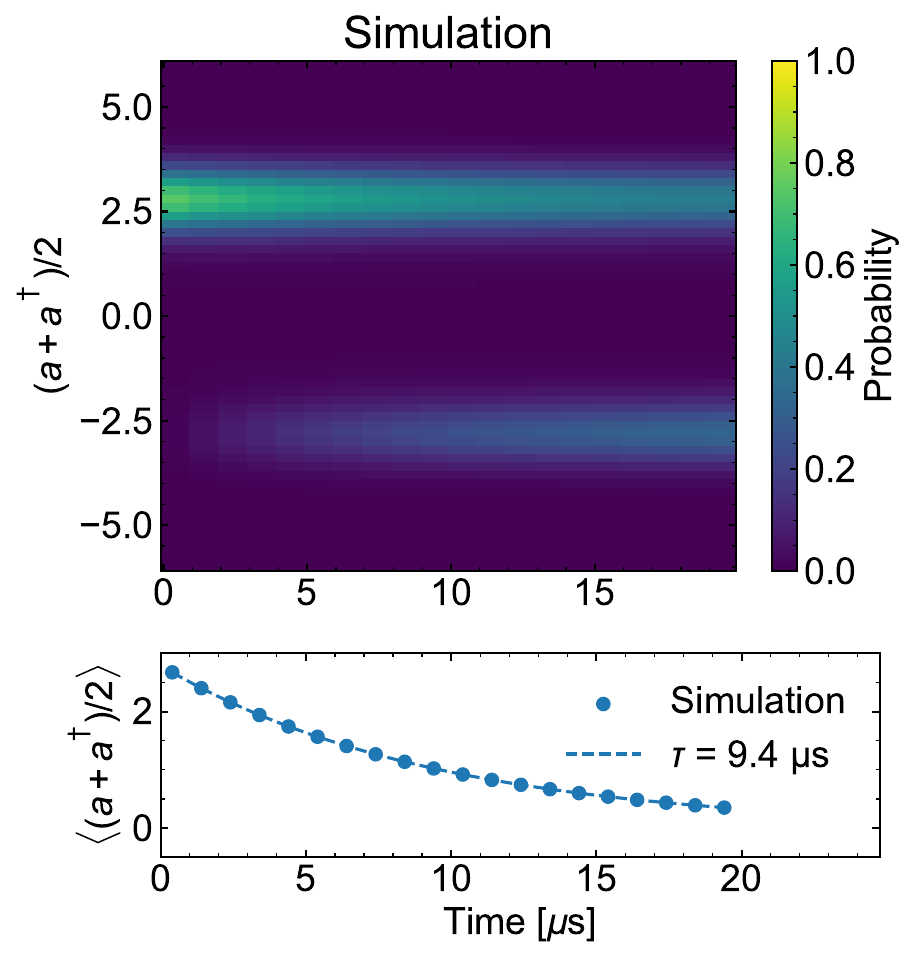}\label{fig:exp1bComparisonSimNoInput}} \\
	\caption{
		(a) Timing chart of the experiment.
		Pump and input signals are pulses with a trapezoidal envelope, where the output voltages of the respective AWGs are increased linearly in 0.2~$\us$ from 0~V to their set values.
		(b) An example histogram of the $IQ$ values of KPO1 without input signal.
		The measurement data for 2.0~$\us$ after the pump pulses have reached the plateau are used.
		A rotation is applied in the $IQ$ plane so that the two peaks are aligned in the horizontal axis, to account for the electrical delay of the pump signal and the readout signal.
		(c) Experimental result of bit-flip time without input signal.
		Upper panel shows the temporal change in the $I$ value of KPO1.
		A condition $I>0$ is applied for the first temporal bin, i.e. $0<t<2~\us$, to select measurement trials with the KPO initially in primarily one of the two qubit states. 
		Lower panel shows the evaluation of the bit-flip time.
		Symbol $\langle I \rangle$ represents the average of $I$ value for each temporal bin.
		The bit-flip time is evaluated using data for both $I>0$ and $I<0$ for the first temporal bin.
		The error of the bit-flip time represents statistical uncertainty.
		(d) Simulation result of bit-flip time without input signal.
		Upper panel shows the temporal change in the probability $\braket{ (a+\ad)/2 | \rho | (a+\ad)/2 }$ averaged over 1~$\us$.
	}
	\label{fig:exp1bMethod}
\end{figure}

The operation parameters are shown in Table~\ref{tab:parameters1b}.
They are evaluated from different experiments as described below.
The resonance frequency $\omegar$, total loss rate $\kappa_\mathrm{tot}$, and external loss rate $\kappa_\mathrm{ext}$ are evaluated by performing a fitting to the scattering parameter of a KPO~\cite{Khalil_2012}.
Here, the external loss rate $\kappa_\mathrm{ext}$ is a coupling between the KPO and the readout line, and the total loss rate $\kappa_\mathrm{tot} \equiv \kappa_\mathrm{ext} + \kappa_\mathrm{int} + 2\gamma$~\cite{Yamaguchi:2023jrp} is the sum of the external loss rate $\kappa_\mathrm{ext}$, internal loss rate $\kappa_\mathrm{int}$, and dephasing rate $\gamma$.
The Kerr nonlinearity $K$ is evaluated by measuring the frequency of the two-photon absorption between the vacuum state $\ket{0}$ and the two-photon Fock state $\ket{2}$~\cite{Yamaji_2022}.
The pump frequency $\omegap$ is set to twice the KPO resonance frequency $\omegar$, so the pump detuning is $\Deltar = \omegar - \omegap/2 = 0~\mathrm{MHz}$.
The pump amplitude $p$ is evaluated from $p=\sqrt{\frac{\Pp}{2Z}}\left|\frac{\mathrm{d}\omegar}{\mathrm{d}I}\right|$~\cite{Yamaguchi:2023jrp}, where $\Pp$ is the pump power, $Z=50~\mathrm{\Omega}$ is the impedance of the pump line, and $\frac{\mathrm{d}\omegar}{\mathrm{d}I}$ is the change in the KPO resonance frequency with respect to the applied flux-bias current.
The pump power $\Pp$ is calibrated based on the excitation energy $\excitation{0}{2}$ as described later in this section.
The coherent-state amplitude $\alpha$ of a KPO in an oscillating state is evaluated from $\alpha = \sqrt{p/|K|}$, which holds when the pump detuning $\Deltar$ is zero~\cite{Goto_2016}.

\begin{table}\centering\caption{Operation parameters of the experiment. Symbol $\Phi_0=h/(2e)$ is the flux quantum.}\label{tab:parameters1b}\begin{ruledtabular}
	\begin{tabular}{ccc}
		                       Parameter                         &             Value                \\ \hline
		                Flux bias $\Phi/\Phi_0 $                 &             0.37              \\
		          Resonance frequency $\omegar/2\pi$ [GHz]            &           9.90            \\
		       Total loss rate $\kappa_\mathrm{tot}/2\pi$ [MHz]        &           1.9             \\
		     External loss rate $\kappa_\mathrm{ext}/2\pi$ [MHz]       &           0.72            \\
		               Kerr nonlinearity $K/2\pi$ [MHz]                &           -14             \\
		Pump detuning $\Deltar/2\pi=(\omegar-\omegap/2)/2\pi$ [MHz] &            0              \\
		                Pump amplitude $p/2\pi$ [MHz]                  & $1.1\times10^2$ \\
		    Coherent-state amplitude $\alpha = \sqrt{p/|K|}$     &             2.8            
	\end{tabular}
\end{ruledtabular}\end{table}

As shown in the timing chart in Fig.~\ref{fig:exp1bTiming}, pump and input signals are injected simultaneously with a trapezoidal envelope whose slope duration is 0.2~$\us$ and plateau duration is 20~$\us$.
They are generated using arbitrary waveform generators (AWGs).
The readout signal from the readout line is sent to amplifiers and a frequency downconverter, then digitized by an analog-to-digital converter (ADC).
Pulse measurements are repeated for many iterations, and the discrete Fourier transform (DFT) is applied to the readout signal to obtain the in-phase and quadrature $IQ$ values as exemplified in the histogram in Fig.~\ref{fig:exp1bIQ}.
The two peaks visible in the histogram correspond to the qubit states of the KPO.
As shown in Fig.~\ref{fig:exp1bComparisonExpNoInput}, the temporal change in the $I$ value is obtained by first dividing the ADC measurement data of each trial into ten temporal bins with a duration of 2~$\us$, then performing the DFT for each bin.
In the top panel of Fig.~\ref{fig:exp1bComparisonExpNoInput}, a condition $I>0$ is applied for the first temporal bin, i.e. $0<t<2~\us$, to select measurement trials with the KPO initially in primarily one of the two qubit states. 
The bit-flip time is evaluated by fitting the temporal change in the average $I$ value with a function $A\exp(-t/\tau)$, where $A$ is a coefficient, $t$ is the time, and $\tau$ is the bit-flip time, as shown in the lower panel of Fig.~\ref{fig:exp1bComparisonExpNoInput}.

A numerical simulation is performed as a comparison.
The time-evolution integration of the Lindblad master equation given below~\cite{Masuda:2024djh} is performed using the simulation software QuTiP~\cite{johansson2012qutip,johansson2013qutip}.
\begin{multline}
	\dot{\rho} = 
		-\frac{i}{\hbar}[H_\mathrm{in}(t),\rho(t)] 
		+ \frac{\kappa_\mathrm{tot}}{2}\left( 2a\rho\ad - \rho \ad a - \ad a\rho  \right) \\
		+ \gamma \left( 2 \ad a \rho \ad a - \rho \ad a \ad a - \ad a \ad a \rho \right). \label{eq:lindblad}
\end{multline}
Here, $\rho$ is the density matrix, $H_\mathrm{in}(t)$ is the Hamiltonian given in Eq.~\eqref{eq:input}, and $\gamma$ is the pure dephasing rate.
The initial state for the numerical simulation is set to the coherent state $\ket{\alpha}$.
Figure~\ref{fig:exp1bComparisonSimNoInput} shows the temporal change in the expectation value $\braket{a+\ad}/2 \equiv \mathrm{Tr}\left[\rho( a+\ad)\right]/2$ calculated from the density matrix $\rho$.
The bit-flip time for the numerical simulation is evaluated by first calculating the expectation value $\braket{a+\ad}/2$ as a function of time, and then performing a fitting with a function $A\exp(-t/\tau)$, as shown in Fig.~\ref{fig:exp1bComparisonSimNoInput}.
The bit-flip time of a KPO decreases as the dephasing rate $\gamma$ increases, because dephasing induces excitation of a KPO~\cite{Masuda:2024djh}, which leads to tunneling and deexcitation that result in bit flips.
The dephasing rate is set to $\gamma/2\pi=9.1~\mathrm{kHz}$ so that the bit-flip time in the simulation approximately matches the experimental result ($\tau=9.4~\mathrm{\mu s}$) when no input signal is applied, as shown in Figs.~\ref{fig:exp1bComparisonExpNoInput} and \ref{fig:exp1bComparisonSimNoInput}.

\begin{figure}
	\centering
	\subfloat[]{\includegraphics[width=.5\linewidth]{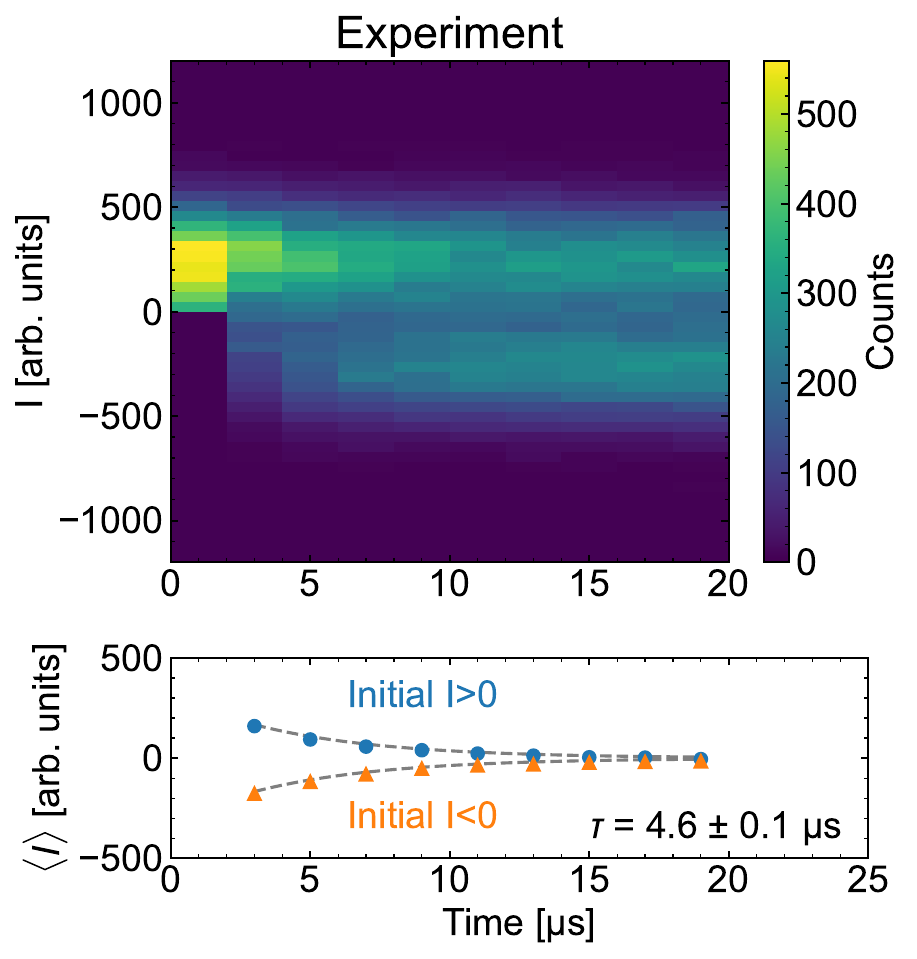}\label{fig:exp1bComparisonExpInput}}
	\subfloat[]{\includegraphics[width=.5\linewidth]{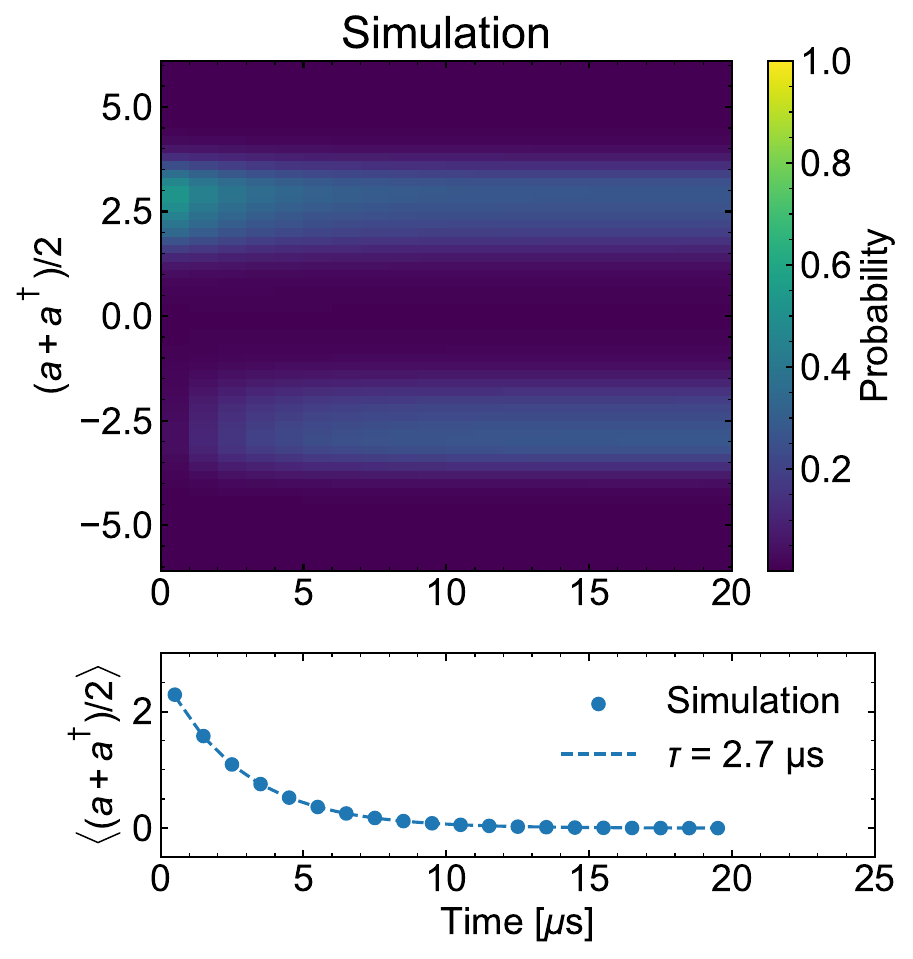}\label{fig:exp1bComparisonSimInput}}
	\caption{
		(a) Experimental and (b) simulation results of the bit-flip time evaluation with an input signal with a detuning $\Deltain/2\pi=-204~\mathrm{MHz}$.
	}
	\label{fig:exp1bComparison}
\end{figure}

The result with an input signal is shown in Fig.~\ref{fig:exp1bComparison}.
The input-signal detuning is $\Deltain=2\pi\times-204~\mathrm{MHz}$, which corresponds to the excitation energy of the KPO among the lowest Hamiltonian eigenstates $\excitation{0}{2}$.
The input-signal amplitude $\Omega_\mathrm{in}$ satisfies the equation below~\cite{Lin_2014}, where $P_\mathrm{in}$ is the input-signal power.
\begin{equation}
	\Omega_\mathrm{in}= \sqrt{ \frac{P_\mathrm{in} \kappa_\mathrm{ext}}{ \hbar \omega_\mathrm{in} } } . \label{eq:inputAmplitude}
\end{equation}
The input-signal power $P_\mathrm{in}$ in the numerical simulation is set to $P_\mathrm{in} = -132~\mathrm{dBm}$, which is based on an independent measurement of the insertion loss of the transmission line, with a 3~dB correction to account for the difference between the experiment and the numerical simulation mentioned above.
This corresponds to $\Omega_\mathrm{in}/2\pi=1.1~\mathrm{MHz}$ based on Eq.~\eqref{eq:inputAmplitude} for $\omega_\mathrm{in}/2\pi=9.90~\mathrm{GHz}$.
As evident from a comparison with Fig.~\ref{fig:exp1bMethod}, an input signal decreases the bit-flip time of KPO1.

\begin{figure}
	\centering
	\subfloat{ \includegraphics[width=.8\linewidth]{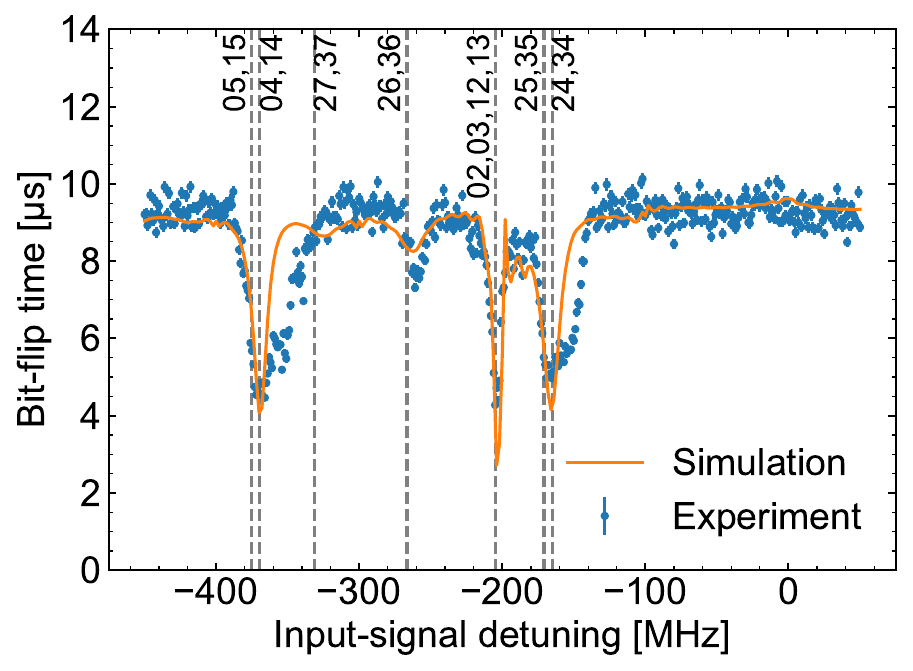}\label{fig:exp1bWrtSim} }
	\caption{
		Bit-flip time of a KPO with respect to $\Deltain$.
		Experimental (blue circle) and simulation (orange line) results are compared.
		Error bars represent statistical uncertainty.
		The results for the input-signal detuning in $-1~\mathrm{MHz}<\Deltain/2\pi<1~\mathrm{MHz}$ are removed, because the readout signal is indistinguishable from the reflection of the input signal and thus cannot be measured.
		The vertical dotted lines represent the KPO excitation energies $\excitation{\idx}{\idy}\equiv\omega_\idy-\omega_\idx$, with the labels representing the indexes $\idx\idy$.
		The eigenenergies $\omega_\idx$ are calculated numerically from the Hamiltonian given in Eq.~\eqref{eq:oneKPO} using the operation parameters shown in Table~\ref{tab:parameters1b}.
	}
	\label{fig:exp1b}
\end{figure}

Figure~\ref{fig:exp1b} displays the bit-flip time with respect to $\Deltain$.
The input-signal power is set to $P_\mathrm{in} = -132~\mathrm{dBm}$ (same as Fig.~\ref{fig:exp1bComparison}).
The pump power $\Pp$ is calibrated by matching the horizontal position of the dip at $\Deltain/2\pi=-204~\mathrm{MHz}$ to the computed excitation energy among the lowest Hamiltonian eigenstates $\excitation{0}{2}/2\pi$, similarly to the method described in Refs.~\cite{Masuda_2021, Yamaguchi:2023jrp}.
This calibration method relies on the relation $|\excitation{0}{2}|\simeq 2p$ for $\alpha\gg 1$, which follows from $|\excitation{0}{2}|\simeq 2|K|\alpha^2$~\cite{Frattini_2024} and $\alpha=\sqrt{p/|K|}$ at $\Deltar=0$~\cite{Goto_2016}.
The data show multiple dips whose horizontal positions correspond to various excitation energies $\excitation{\idx}{\idy}$ between Hamiltonian eigenstates.
Resonant excitation ($\Deltain= \excitation{\idx}{\idy}$) followed by tunneling and deexcitation results in bit flips, as described in Sec.~\ref{sec:bitflip}.

\begin{figure*}
	\centering
	\subfloat[]{\includegraphics[height=4cm]{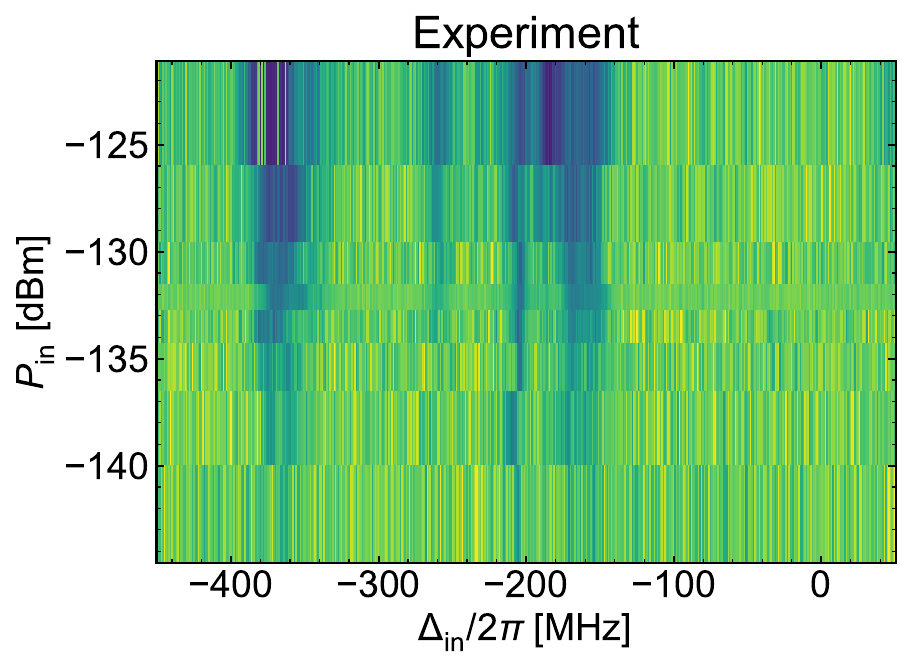}\label{fig:exp1bVsPower}}
	\subfloat[]{\includegraphics[height=4cm]{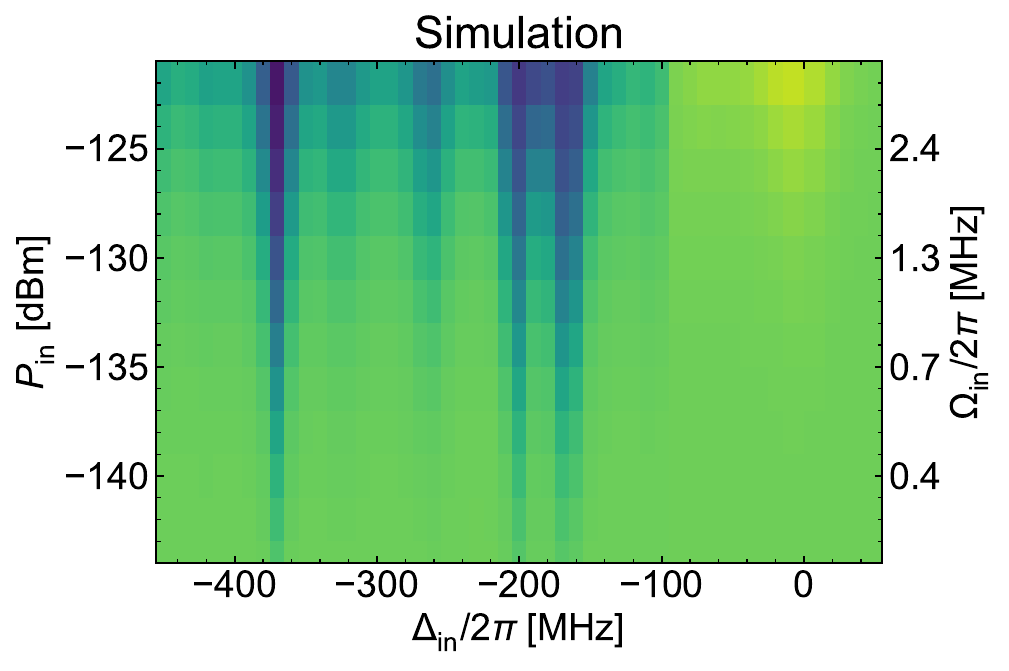}\label{fig:sim1bVsPower}}
	\subfloat[]{\includegraphics[height=4cm]{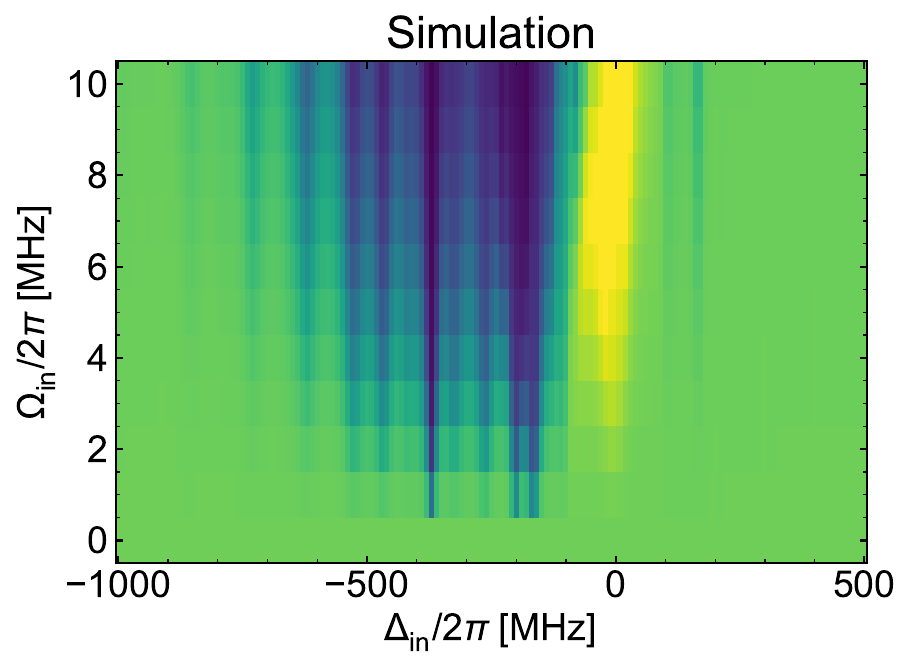}\label{fig:sim1bVsAmplitude}}
	\subfloat{\includegraphics[height=4cm]{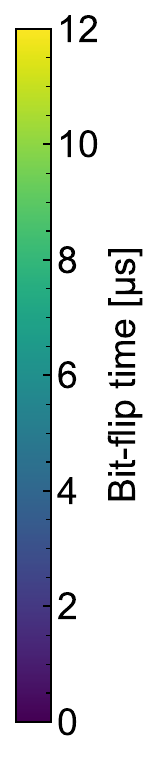}}
	\caption{
		(a), (b), Bit-flip time with respect to the input-signal detuning and the input-signal power, for experiments and numerical simulations respectively.
		The results for the input-signal detuning in $-1~\mathrm{MHz}<\Deltain/2\pi<1~\mathrm{MHz}$ are removed, because the readout signal is indistinguishable from the reflection of the input signal and thus cannot be measured.
		In (a), the statistical fluctuations of the bit-flip times are smaller for the row corresponding to the input-signal power $P_\mathrm{in}=-132~\mathrm{dBm}$ compared to those of other rows,
		because $10^4$ trials of pulse measurements are performed for each value of the input-signal detuning $\Deltain$ for input-signal power $P_\mathrm{in}=-132~\mathrm{dBm}$, and $10^3$ trials are performed for each $\Deltain$ for other values of the input-signal power.
		In (b), for the vertical axis, values of the input-signal amplitude $\Omega_\mathrm{in}$ corresponding to those of the input-signal power $P_\mathrm{in}$ for input-signal frequency $\omega_\mathrm{in}/2\pi=9.90~\mathrm{GHz}$ are also shown.
		(c) Simulation result with wider ranges in the horizontal and vertical axes.
	}
	\label{fig:exp1bdeltavspowervslifetime}
\end{figure*}

Figures~\ref{fig:exp1bVsPower} and \ref{fig:sim1bVsPower} show the dependence of bit-flip time on $\Deltain$ and $P_\mathrm{in}$ for the experiment and numerical simulation, respectively.
Similarly to Fig.~\ref{fig:exp1b}, results show multiple dips in the horizontal axis with the horizontal positions corresponding to various $\excitation{\idx}{\idy}$.
The decrease in bit-flip time becomes more significant with higher input-signal power, because higher input-signal power increases the excitation rate.
Results in Figs.~\ref{fig:exp1bVsPower} and \ref{fig:sim1bVsPower} are in qualitative agreement.

Figure~\ref{fig:sim1bVsAmplitude} shows the simulation result with wider ranges in the horizontal and vertical axes, to provide more insight into the dependence of bit-flip time on $\Deltain$ and $\Omegain$.
An increase in bit-flip time is observed when $|\Deltain|/2\pi \lesssim 100~\mathrm{MHz}$.
This arises because an input signal that is on-resonant ($\Deltain=0$) tilts the metapotential~\cite{Lin_2014}, thus increasing the well depth of one of the two wells and decreasing that of the other, depending on $\thetain$.
For example, based on a numerical simulation when the initial state is $\ket{\alpha}$, $\Deltain=0$, and $\Omegain/2\pi=3~\mathrm{MHz}$, the bit-flip time is increased to 15~$\us$ when $\thetain=0$, while it is 10~$\us$ when $\thetain=\pi$ and remains approximately unchanged compared to when the input signal is absent.
The increase when $\thetain=0$ is because of the increase in the well depth.
However, the lack of decrease when $\thetain=\pi$ is not understood, and may be related to the nonlinear relationship between the well depth and the bit-flip time described in Ref.~\cite{Su_2025}.
For non-resonant but small $|\Deltain|$ as in Fig.~\ref{fig:sim1bVsAmplitude}, the input signal can be effectively viewed as an input signal with $\Deltain=0$ and a rotating phase, resulting in an averaging of the effect with respect to $\thetain$ and an increase in the bit-flip time.

	\section{Discussion}\label{sec:discussion}

From the discussions in Secs.~\ref{sec:bitflip} and \ref{sec:exp1b}, we infer that in a multi-KPO system, the bit-flip time of a KPO can potentially decrease when half of the difference between
pump frequencies of KPOs matches an excitation energy of a KPO, $\Deltap = \excitation{\idx}{\idy}$.
The degradation becomes stronger with larger coupling-amplitude product $g\alpha$.

To mitigate this degradation in multi-KPO systems, two complementary strategies can be employed.
One approach is to adjust the parameters of KPOs so that the relationship $\Deltap \neq \excitation{\idx}{\idy} (\idx<\idy)$ holds.
\begin{itemize}
    \item The left-hand side $\Deltap = (\omegap_2-\omegap_1)/2$ can be adjusted by changing the pump frequencies $\omegap$ of KPOs.
    The frequency configuration has to be chosen carefully considering the architecture of the chip, for instance that described in Ref.~\cite{Puri_2017}.
    \item The right-hand side $\excitation{\idx}{\idy}$ can be adjusted by changing the parameters of the Hamiltonian in Eq.~\eqref{eq:oneKPO}, for instance the pump detuning $\Deltar$, the Kerr nonlinearity $K$, and the pump amplitude $p$.
\end{itemize}

Another approach is to decrease the product $g\alpha$.
\begin{itemize}
    \item The two-body coupling $g$ between KPOs can be decreased by changing the design of the qubit chip.
    For example, for intentional couplings by design, tunable couplers can be employed~\cite{Aoki_2024}.
    For unintentional capacitive couplings between KPOs, through-silicon vias or airbridges can be used to decrease the coupling.
    For effective coupling via a common readout line for frequency-multiplexed readout, the relative placement among KPOs along the readout line can be optimized~\cite{van_Loo_2013,Lalumi_re_2013,Gheeraert_2020}.
    Experimental demonstration of the frequency-multiplexed readout of KPOs is described in Appendix~\ref{sec:exp2b}.
    \item The coherent-state amplitude $\alpha$ can be decreased, however with a possible decrease in measurement sensitivity~\cite{Yamaji_2022}.
\end{itemize}
	\section{Conclusion}
We have experimentally characterized how interactions between KPOs degrade bit-flip times in multi-KPO quantum systems.
Photon exchanges caused by interactions between KPOs induce excitations, followed by tunneling and deexcitation that lead to bit flips.
Thus, inter-KPO interactions can lead to a decrease in bit-flip times of KPOs.

In an experiment in which a weak microwave signal is injected into a KPO to emulate photon injection from another KPO, we have observed that the bit-flip time of the KPO decreases by an order of magnitude when the input-signal detuning matches the excitation energy of the KPO.
The degradation increases with input-signal power.

Methods to mitigate the decrease in bit-flip times caused by interactions between KPOs have been discussed, for instance  adjusting the pump frequencies, coherent-state amplitudes, and couplings between KPOs.
These findings provide valuable insights for scaling KPO-based quantum computers to many-qubit systems while maintaining high fidelity.
	\begin{acknowledgments}
	
We thank Y. Matsuzaki, T. Ishikawa, K. Inomata, and K. Mizuno for fruitful discussions.
The KPO chip was fabricated in the Superconducting Quantum Circuit Fabrication Facility (Qufab) in National Institute of Advanced Industrial Science and Technology (AIST).
This paper is based on results obtained from a project, JPNP16007, commissioned by the New Energy and Industrial Technology Development Organization (NEDO).

\end{acknowledgments}
	\appendix
	\section{Experimental setup}\label{sec:setup}

\begin{figure*}
	\centering
	\includegraphics[width=1\linewidth]{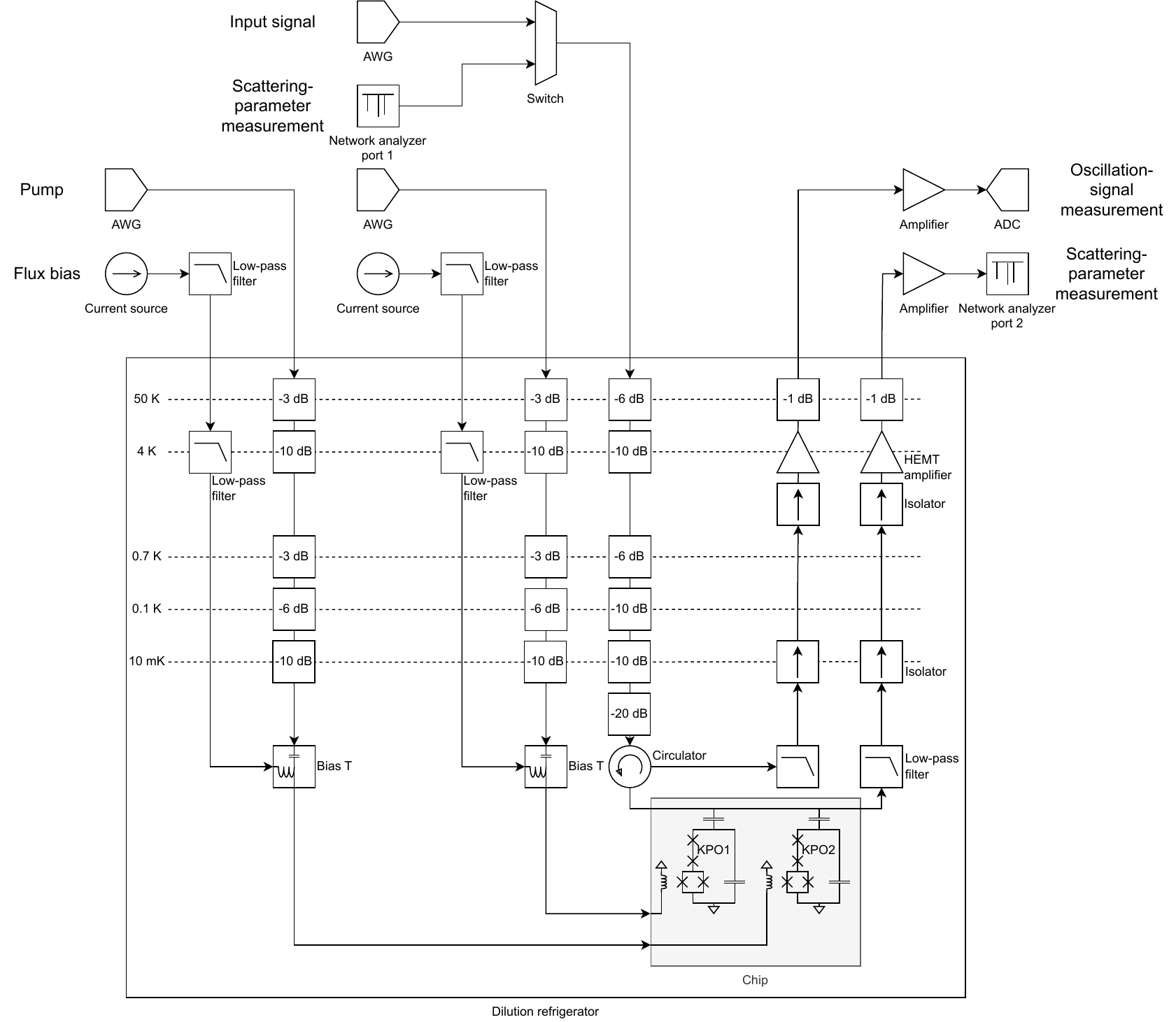}
	\caption{
			Schematic diagram of the experimental setup.
			Components that have low insertion loss and are irrelevant to the results (e.g. switches) are omitted for simplicity.
			Boxes with insertion losses written inside represent attenuators.
		}
	\label{fig:experimentalsetup}
\end{figure*}

The experimental setup is shown in Fig.~\ref{fig:experimentalsetup}.
Instruments placed in a room-temperature environment and connected to the dilution refrigerator generate the signals injected into KPOs (e.g. pump signal, flux-bias current, and input signal), and also process the output signals from KPOs.

	\section{Frequency-multiplexed readout of two KPOs}\label{sec:exp2b}

\begin{figure}
	\centering
	\subfloat[]{\includegraphics[width=.5\linewidth]{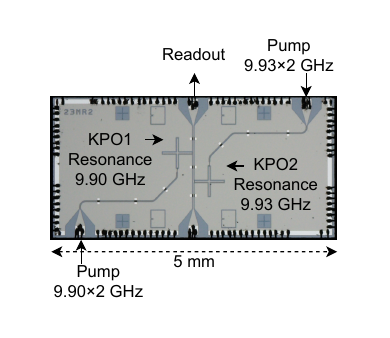}\label{fig:setup2b}}
	\subfloat[]{\includegraphics[width=.5\linewidth]{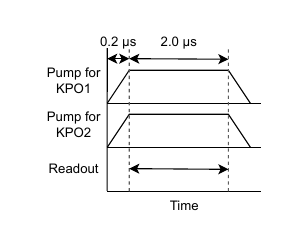}\label{fig:exp2bPulseChart}}
	\caption{
		(a) Experimental setup for the demonstration with two KPOs.
		(b) Timing chart of the experiment. 
		}
	\end{figure}

We describe the experimental demonstration of frequency-multiplexed readout of KPOs.
As shown in Fig.~\ref{fig:setup2b}, pump signals are applied simultaneously to both KPOs to induce their oscillations, and the two oscillation signals are read out simultaneously from the readout line.
As shown in the timing chart in Fig.~\ref{fig:exp2bPulseChart}, both pump pulses are simultaneously injected into the two KPOs.
An input signal is not applied.

The operation parameters of the two KPOs in this experiment are described in Table~\ref{tab:parameters2b}.
Those for KPO1 match those in Table~\ref{tab:parameters1b}, except for slight adjustments in pump amplitude $p$ and the resulting coherent-state amplitude $\alpha$.
The coupling $g$ between the two KPOs is determined from a measurement of the avoided crossing, where the flux-bias current of KPO2 is varied and the scattering parameter of the two-KPO system is measured with the setup shown in Fig.~\ref{fig:setup2b}.
The minimum frequency separation of the two resonance dips is $2g/2\pi=6.0~\mathrm{MHz}$, which agrees with electromagnetic-field simulation based on the design of the KPO chip.
The coupling is understood to be dominated by capacitive coupling between the two KPOs, because approximately the same values are obtained from the electromagnetic-field simulation described above and from another simulation based on the same design excluding the readout line.
Also, the effective coupling via the readout line is estimated to be less than 0.1~MHz based on Ref.~\cite{Gheeraert_2020}.

\begin{table*}\centering\caption{Operation parameters of the two KPOs.}\label{tab:parameters2b}\begin{ruledtabular}
	\begin{tabular}{ccc}
		                       Parameter                         &             KPO1             &             KPO2             \\ \hline
		                Flux bias $\Phi/\Phi_0 $                 &             0.37             &             0.36             \\
		          Resonance frequency $\omegar/2\pi$ [GHz]           &           9.90            &           9.93            \\
		       Total loss rate $\kappa_\mathrm{tot}/2\pi$ [MHz]        &           1.9             &           2.1             \\
		     External loss rate $\kappa_\mathrm{ext}/2\pi$ [MHz]       &           0.72            &           0.64            \\
		               Kerr nonlinearity $K/2\pi$ [MHz]                &           -14             &           -15             \\
		Pump detuning $\Deltar/2\pi=(\omegar-\omegap/2)/2\pi$ [MHz] &            0              &            0              \\
		                Pump amplitude $p/2\pi$ [MHz]                  & $1.4\times10^2$ & $1.4\times10^2$ \\
		    Coherent-state amplitude $\alpha = \sqrt{p/|K|}$     &             3.1              &             3.0              \\
		               Two-body coupling $g/2\pi$ [MHz]                &                 \multicolumn{2}{c}{3.0}
	\end{tabular}
\end{ruledtabular}\end{table*}

\begin{figure}
	\centering
	\subfloat[]{\includegraphics[width=.8\linewidth]{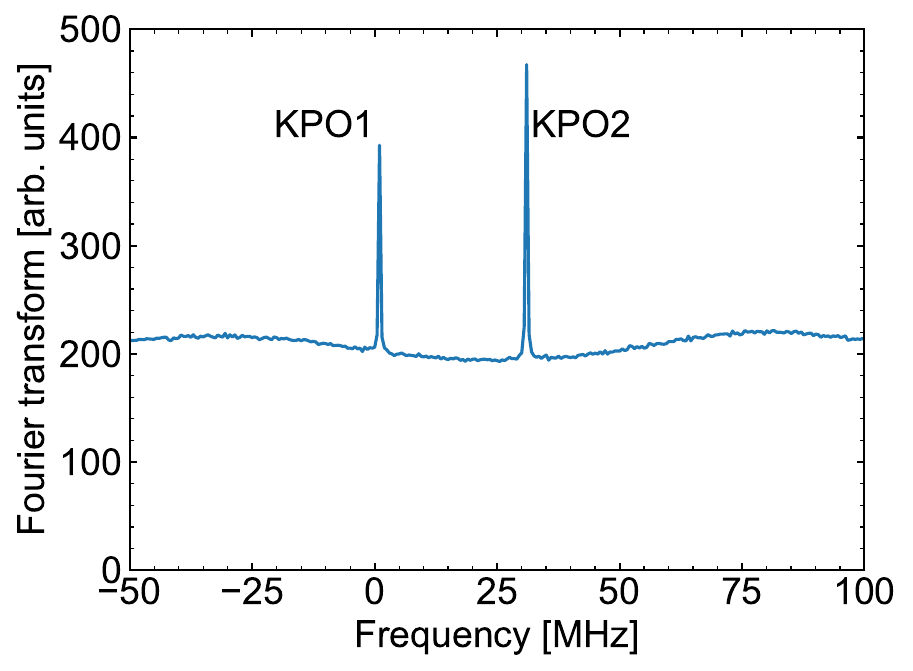}\label{fig:exp2bFourier}}\\
	\subfloat[]{\includegraphics[width=\linewidth]{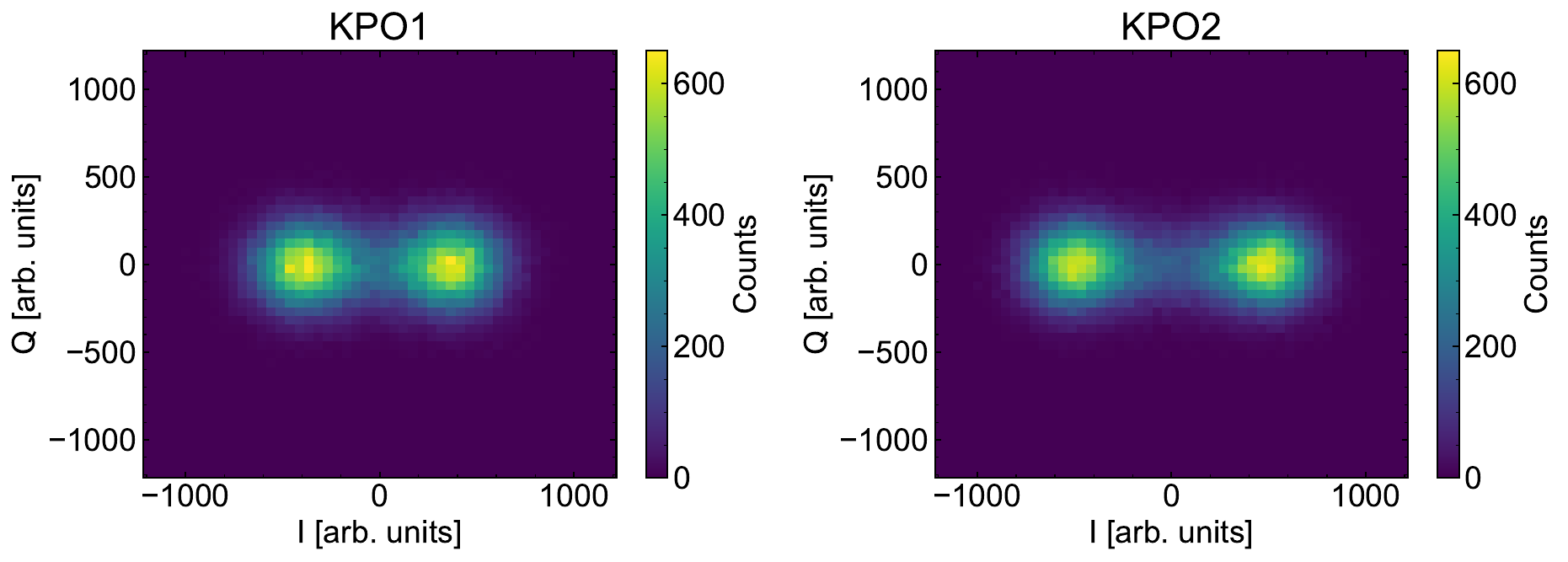}\label{fig:exp2bIQ}}
	\\
	\subfloat[]{\includegraphics[width=.8\linewidth]{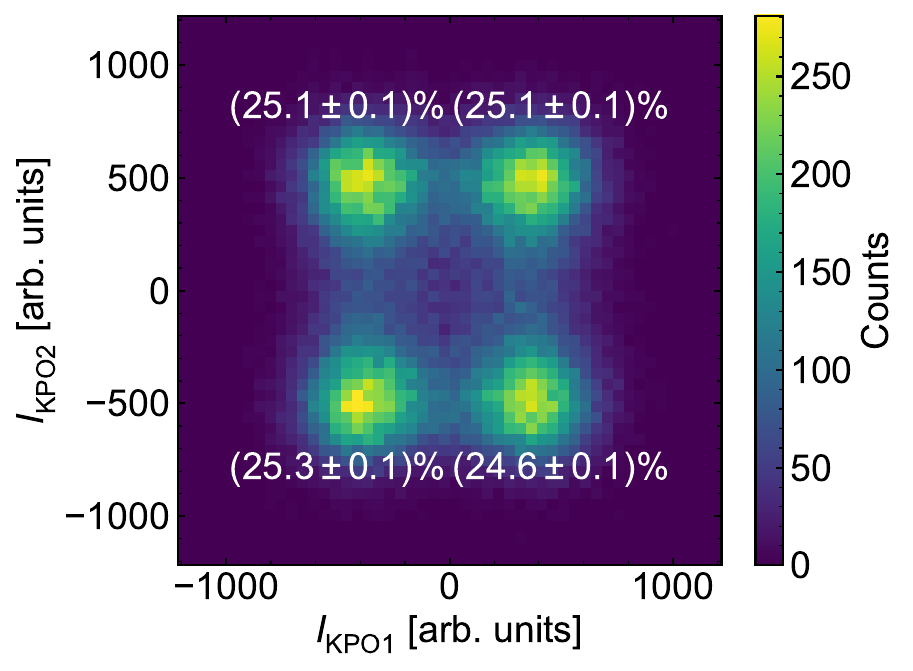}\label{fig:exp2bCorrelation}}
	\caption{
		Frequency-multiplexed readout of two KPOs.
		(a) DFT spectrum of the measurement data of an ADC, averaged over $10^4$ trials of pulse measurements.
		(b) Histogram of the $IQ$ values of the two KPO oscillation signals.
		(c) Histogram showing the $I$ values of KPO1 $I_\mathrm{KPO1}$ and KPO2 $I_\mathrm{KPO2}$. 
		Also shown are the ratios of measurement trials in four quadrants. 
		Errors indicate the statistical uncertainties.
	}
	\label{fig:exp2b}
\end{figure}

Figure~\ref{fig:exp2bFourier} shows the DFT spectrum of the measurement data of an ADC.
The output signals of the two KPOs at 9.90~GHz and 9.93~GHz are down converted to 1~MHz and 31~MHz, respectively, by a frequency downconverter prior to ADC measurement, and thus the two peaks at 1~MHz and 31~MHz in Fig.~\ref{fig:exp2bFourier} correspond to the two signals.
Figure~\ref{fig:exp2bIQ} shows histograms of the DFT values plotted in $IQ$ planes at 1~MHz and 31~MHz, corresponding to KPO1 and KPO2 respectively.
Figure~\ref{fig:exp2bCorrelation} presents the histogram of $I$ values for both KPOs.
The ratios of the observed counts in the four quadrants are approximately 25\%, indicating that the qubit states of the two KPOs are uncorrelated, as expected because the pump frequencies of the two KPOs are different~\cite{Yamaji_2023}.

The method described here can be used to investigate the mitigation of decrease in bit-flip times described in Sec.~\ref{sec:discussion} in the context of frequency-multiplexed readout.
For instance, one can design a multi-KPO system with the consideration in Sec.~\ref{sec:discussion} and evaluate the bit-flip times of KPOs using frequency-multiplexed readout.
	\bibliography{refs.bib}
\end{document}